\documentclass{amia}
\usepackage{hyperref}
\usepackage{booktabs}
\usepackage{amsfonts}
\usepackage{amssymb}
\usepackage{amsthm}
\usepackage{amsmath}
\usepackage{nicefrac}
\usepackage{multirow}
\usepackage{subfigure}
\usepackage{diagbox}
\usepackage{algpseudocode}
\usepackage{algorithm}
\usepackage[framemethod=tikz]{mdframed}
\usepackage{tcolorbox}
\usepackage{adjustbox}
\usepackage{wrapfig}

\begin{document}

\title{PRIME: Uncovering Circadian Oscillation Patterns and Associations with AD in Untimed Genome-wide Gene Expression across Multiple Brain Regions}

\author{Xinxing Wu, PhD$^a$, Chong Peng, PhD$^b$, Gregory Jicha, PhD, MD$^a$, Donna Wilcock, PhD$^a$, Qiang Cheng, PhD$^{a,}${\footnote{Corresponding should be sent to: qiang.cheng@uky.edu.}}}
\institutes{
    $^a$University of Kentucky, Lexington, Kentucky, USA; $^b$Qingdao University, Qingdao, Shandong, China
}

\maketitle

\section*{Abstract}
\vskip -0.1in

\textit{The disruption of circadian rhythm is a cardinal symptom for Alzheimer’s disease (AD) patients. The full circadian rhythm orchestration of gene expression in the human brain and its inherent associations with AD remain largely unknown. We present a novel comprehensive approach, PRIME, to detect and analyze rhythmic oscillation patterns in untimed high-dimensional gene expression data across multiple datasets. To demonstrate the utility of PRIME, firstly, we validate it by a time course expression dataset from mouse liver as a cross-species and cross-organ validation. Then, we apply it to study oscillation patterns in untimed genome-wide gene expression from 19 human brain regions of controls and AD patients. Our findings reveal clear, synchronized oscillation patterns in 15 pairs of brain regions of control, while these oscillation patterns either disappear or dim for AD. It is worth noting that PRIME discovers the circadian rhythmic patterns without requiring the sample’s timestamps. The codes for PRIME, along with codes to reproduce the figures in this paper, are available at https://github.com/xinxingwu-uk/PRIME.}

\section*{Introduction}
\vskip -0.1in

Alzheimer’s disease (AD) is the leading form of dementia. Unlike other chronic diseases such as heart disease and cancer, it has an increasing death rate worldwide. Multiple studies have shown that circadian rhythm and AD have a two-way relationship, and most circadian disruptions typically occur in the early stage of AD, even preceding the development of cognitive symptoms~\citep{Gagnon,Yue,Irwin,Noble}. The circadian clock system is endogenously generated by an internal circadian biological clock~\citep{Jay,David}. In mammals, circadian rhythm is governed by a brain region called {\sl suprachiasmatic nucleus (SCN)}, a group of cells in the hypothalamus that respond to light and dark signals. There is growing evidence showing the circadian clock controls behavioral phenotypic rhythms and harmonizes peripheral clocks located in almost every organ, such as the brain, liver, kidney, and heart~\citep{Ueli,Michael,Akhilesh,Dibner}. A plethora of clinical, molecular, and genomic studies analyze the pathophysiology of AD and its associations with circadian disorders~\citep{Susanna,Lucey,Jun,Cho}, and accumulating evidence indicates that the disorder of circadian rhythm is not only a pathological marker but a putative risk factor of AD. Currently, a clear understanding of the relationships between the disturbances of the human brain's circadian rhythms and AD from genome-wide gene expression remains elusive. 

Existing studies on the relationships between circadian rhythm and AD mainly focus on a limited number of single clock genes, for instance, {\sl Bmal1} (also known as {\sl ARNTL}), {\sl Cry1}, {\sl Per1}, and {\sl Per2}~\citep{Ying,Nicolas,Hyundong}, or brain regions such as pineal gland and cingulate cortex and bed-nucleus of the stria terminalis~\citep{Nicolas}. Almost all these studies are based on mouse models, while little is known regarding the impact on humans. More intriguingly, these prior analyses of circadian rhythm all depend on time labels of samples, such as time-of-death, while such time labels are practically difficult or inaccurate, if not impossible, to obtain through clinically interval sampling for human brain tissue due to the highly invasive nature. As pointed out in~\citep{Ron}, more than 1 million human gene expression samples in the National Center for Biotechnology Information Gene Expression Omnibus (GEO) repository are available for discovery, but these samples' collection time was rarely recorded. Recently, two computational approaches have been developed that allow an analysis of a single clock gene's circadian rhythms of untimed sample data~\citep{Ron,Ning}. Such single gene-based studies can facilitate our understanding of the principles of the biological clock at the gene level. However, the circadian clock and peripheral clocks are macroscopically reflected at the tissue and organ level for organisms. Also, different clock genes may have much varied circadian expression levels. For example, {\sl Bmal1} and {\sl Per1} display antiphase circadian expressions~\citep{Caldelas}; {\sl ARNTL} and {\sl Clock} present different seasonality expressions~\citep{Xaquin}. Because the circadian and peripheral clocks are the overall orchestration of the circadian rhythms or oscillation patterns of many clock genes that comprise these tissues and organs, it is overly simplistic to take the single clock genes' circadian rhythms as a direct characterization of the tissues and organs. In the literature, another research line is to study oscillation patterns for untimed samples at a single cell level, e.g., the studies in~\citep{Gut,Yusen}. Nonetheless, the core of their studies along such a research line is mainly to analyze the periodicity of cell cycles rather than circadian patterns at the tissue level. 

The orchestration of circadian oscillation patterns of all relevant genes of the human brain and the potential dysregulation impact on AD remain unknown; there is an urgent need to develop methods to analyze whole-genome gene expression data to find circadian oscillation patterns of different brain regions. One main barrier to studying circadian patterns of genome-wide gene expression is the requirement of time labels of brain tissues, which are hard to obtain or inaccurate. This paper explores the relationships between circadian rhythms of brain regions and AD with genome-wide gene expression data. We present an innovative, comprehensive approach based on principal component analysis (PCA) and matrix factorization (MF), called PRIME, to detect and analyze oscillation patterns in untimed high-dimensional data across multiple brain regions. To validate PRIME, we first apply it to a time course expression dataset from mouse livers. Then, we use it to analyze the connections of AD with circadian rhythms based on a dozen of large-scale gene expression datasets from 19 cortical regions of about 60 human subjects. Furthermore, we comprehensively evaluate and validate our discovery using statistical correlation analyses, such as Pearson, Spearman, and Kendall correlations, on 8 well-known circadian clock genes, i.e., {\sl ARNTL}, {\sl ACSNK1E}, {\sl ANPAS2}, {\sl ANR1D2}, {\sl APER1}, {\sl APER2}, {\sl APER3}, and {\sl ARORA}~\citep{Jun, Cho,Morales} in various brain regions. Our study reveals the genome-wide gene expression-based correlations between circadian disorders in different brain regions and AD: There exist clear, synchronized oscillation patterns in 15 pairs of brain regions of control, including middle temporal gyrus and inferior temporal gyrus, frontal pole and anterior cingulate, frontal pole and dorsolateral prefrontal cortex, hippocampus and dorsolateral prefrontal cortex, and hippocampus and prefrontal cortex. Notably, these circadian rhythmic oscillation patterns either disappear or dim for AD.

Our approach consists of four steps: Firstly, it merges cross-region datasets from 19 brain regions such as the middle temporal gyrus, inferior temporal gyrus, and frontal pole; Secondly, by doing sample cross-reference on union brain regions, it obtains the overlap of samples on different brain regions, see Figure~\ref{fig:01} (a); Thirdly, it uses PRIME listed in {\bf Algorithm~\ref{alg:pm}} to obtain circadian rhythmic oscillation patterns; Finally, it computes the Pearson, Spearman, and Kendall correlation coefficients for controls and AD patients on different brain regions. According to whether labels are used or not, circadian rhythm-related analysis can be classified into supervised, such as ZeitZeiger~\citep{Jacob} which is used to estimate a periodic variable from high-dimensional observations, and unsupervised ~\citep{Ron,Ning,Gut}. Our proposed approach belongs to the latter, with no need for massive labeled samples to do training. {\sl To the best of our knowledge}, it is the first systematic study to investigate the associations between multiple brain regions' circadian rhythms and AD quantitatively based on untimed genome-wide gene expression, which permits the most direct glimpse into the circadian events in the human brain at the brain region- or tissue-level. %The codes for PRIME, along with codes to reproduce the figures in this paper, will be made publicly available on GitHub upon acceptance.

\section*{Methods}
\vskip -0.1in

%Next, we present the mathematical and algorithmic descriptions of PRIME and its implementation details.
%------------------------------------------------------------------------------------------------------------------------------------------------------
\paragraph{Notations}

Let $N$ and $M$ be the numbers of samples and probes, respectively. Denote $\mathbf{x}_{m(1:N)}=\left[x_{m1}, \ldots, x_{mN}\right]$, $m=1,\ldots,M$; let $\mathbf{X}=\left[\mathbf{x}_{1(1:N)}^T,\ldots,\mathbf{x}_{M(1:N)}^T\right]^T\in\mathbb{R}^{M\times N}$ be the sample matrix, and $\mathbf{x}_{(1:M)n}=\left[x_{1n}, \ldots, x_{Mn}\right]^T$ be the genome-wide gene expression of a sample, with $x_{mn}$ denoting the expression value of probe $m$ for sample $n$. For each sample, its label is known as either control or AD.

%------------------------------------------------------------------------------------------------------------------------------------------------------
\paragraph{Data normalization}

The gene expression $x_{mn}$ for each probe $m$ of sample $n$ is normalized as ${\overline{x}_{mn}}=({\displaystyle x_{mn}-\mu_m})/$ $({\sqrt{\frac{1}{N}\sum_{n=1}^N(x_{mn}-\mu_m)^2}})$, where $\mu_m$ is the averaged expression of probe $m$ across all samples, $\mu_m=\frac{1}{N}\sum_{n=1}^N{x_{mn}}$, $m=1,\ldots,M$. We denote ${\overline{\mathbf{x}}_{m(1:N)}}=\left[{\overline{x}_{m1}}, \ldots, {\overline{x}_{mN}}\right]$.

\paragraph{PRIME for capturing oscillatory patterns in the latent space}

\begin{figure}[!htbp]
\vskip -0.14in
  \centering
    \includegraphics[width=1\textwidth]{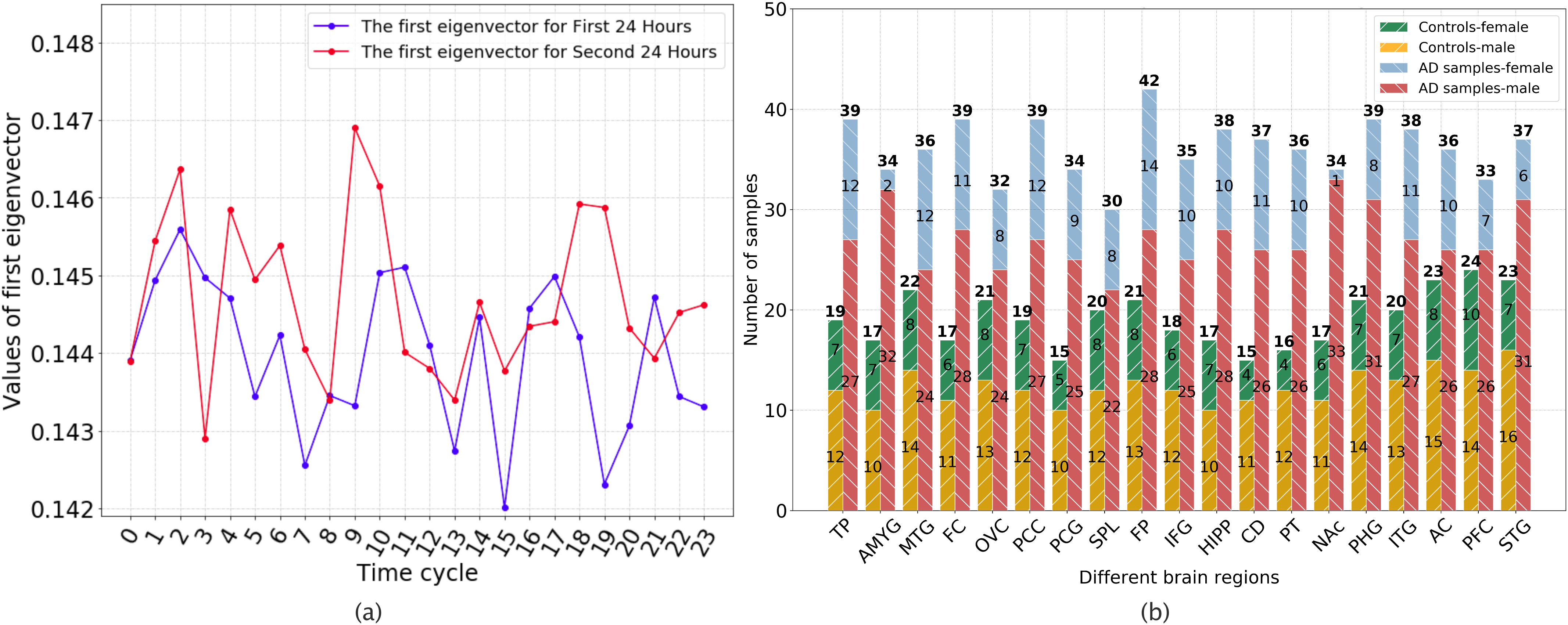}
  \vskip -0.14in
 \caption{(a) The first eigenvectors of mouse liver data in two cycles. The horizontal axis denotes the order number of samples in a cycle. We calculated that their Pearson {\sl correlation = 0.2, p = 0.381}, Spearman {\sl correlation = 0.2, p = 0.355}, and Kendall {\sl correlation = 0.2, p = 0.298}. Although the samples are from two consecutive periodic cycles, the result of eigengenes shows they are uncorrelated. Thus, the \lq\lq eigengene\rq\rq\,\, method fails to work for this data. (b) Distributions with respect to controls and AD samples in men and women. The vertical axis represents the number of samples, with more AD samples than the control samples.}
 \label{fig:01}
\end{figure}

By applying the singular value decomposition (SVD) on the gene expression data, \citep{Alter} obtained the \lq\lq eigengenes," which are the eigenvectors corresponding to the singular values. \citep{Alter} also pointed out that the first \lq\lq eigengene\rq\rq\,\,could be used to obtain out-of-phase sinusoidal oscillations. All these first \lq\lq eigengenes\rq\rq\,\,with the samples' collection times construct an ellipse~\citep{Ron}.{\footnote{Note that \citep{Ron} appears loose to use a circular neural network to extract angles or phases, because a circle implies a uniform velocity, whereas an ellipse implies a non-uniform velocity.}} However, the genome-wide expression data is high-dimensional, comprising a large number of genes that are essentially irrelevant to circadian rhythms and can be regarded as noise. Therefore, the direct use of\,\,\lq\lq eigengenes\rq\rq\,\,is strongly susceptible to noise and possible corruption in the data.{\footnote{Taking the mouse liver dataset as an example, we compute and plot the first \lq\lq eigengenes\rq\rq\,\, from two entire periodic cycles in Figure~\ref{fig:01} (a).}} To have an effective algorithm resilient to irrelevant genes and noise, we will capture the circadian rhythmic patterns in the latent space underlying the gene expression data. Our technique first relies on PCA to reduce the dimension of the gene expression data, followed by applying MF to reveal the latent groups in an intrinsic low-dimensional subspace and the hidden relationship between these latent groups and the subjects. Finally, PRIME extracts potential oscillatory patterns in the resultant latent space. We describe the methodological details in three main steps below:

{\bf Step 1}. Applying PCA to the normalized data, we obtain the dimension-reduced data $\overline{\mathbf{X}}^{D}=[\mathbf{x}_{(1:D)1},\ldots,\mathbf{x}_{(1:D)N}]$, where $D\ll M$. Then we utilize the singular value decomposition of the normalized data matrix to compute PCA.

{\bf Step 2}. Using MF, we decompose $\overline{\mathbf{X}}^{D}$ as follows:
\vspace{-1em}
\begin{equation}{\label{facEqu1}}
\displaystyle\min_{\mathbf{V},\mathbf{G}}\left\|\overline{\mathbf{X}}^{D}-\mathbf{V}\cdot\mathbf{G}^T\right\|_{\mathrm{F}}^2,
\end{equation}
\vspace{-2em}

where $\mathbf{V}\in\mathbb{R}^{D\times R}$, and $\mathbf{G}\in\mathbb{R}^{N\times R}$, with $R$ being a pre-specified parameter for the latent space's dimension. $\mathbf{V}\cdot\mathbf{G}^T$ represents the multiplication of $\mathbf{V}$ and $\mathbf{G}^T$. $\|\cdot\|_{\mathrm{F}}$ is the Frobenius norm, $\|\mathbf{X}\|_{\mathrm{F}}=\sqrt{\sum_{m=1}^M\sum_{n=1}^N|x_{mn}|^2}$. Generally speaking, $\mathbf{V}$ comprises the representatives of $R$ latent clusters in the space spanned by $\overline{\mathbf{X}}^{D}$, and $\mathbf{G}$ represents the (soft) membership metrics of all samples on $R$ clusters in the latent space~\citep{Chris}.

{\bf Step 3}. Performing enhancement of the membership metrics. Denoting $\odot$ as the Hadamard product of matrices, we compute $\mathbf{G}\odot\mathbf{G}$ as our result containing the oscillatory patterns in the latent space. This operation is analogous to the contrast enhancement performed in image processing~\citep{Pratt}. We denote $\mathbf{G}\odot\mathbf{G}$ by $\mathbf{G}^2$ for notational simplicity. 

The overall algorithm consists of the above three steps, which we summarize in {\bf Algorithm~\ref{alg:pm}}.

%------------------------------------------------------------------------------------------------------------------------------------------------------
\begin{algorithm}[!htbp]
  \caption{PRIME.}
  \label{alg:pm}
  \footnotesize
  \begin{algorithmic}[1]
    \Require
      A dataset from a brain region $\mathbf{X}_{M\times N}$; A percentage threshold  $\beta$
    \Ensure
      Cluster (soft) membership metrics matrix $\mathbf{G}^2$
    \State Normalize $\mathbf{X}_{M\times N}$;
    \State Compute singular values $\lambda_i$ and right-singular vectors $\overline{\mathbf{x}}_{i(1:N)}, i=1,2,\ldots, M$;
    \State Retain the top $D$ eigenvectors determined by the percentage threshold $\min\limits_{D}{\sum_{i=1}^{D}\lambda_i}/{\sum_{j=1}^{M}\lambda_j}\geqslant\beta$;
    \State Let $\overline{\mathbf{X}}\in\mathbb{R}^{D\times N}$ be the matrix containing vectors $\overline{\mathbf{x}}_{(1:D)n}, n=1,2,\ldots,N$;
	\State Do MF on $\overline{\mathbf{X}}$ by~(\ref{facEqu1}) to obtain $\mathbf{G}$;
	\State Perform contrast enhancement to obtain $\mathbf{G}^2$;\\
    \Return $\mathbf{G}^2$
  \end{algorithmic}
\end{algorithm}
%------------------------------------------------------------------------------------------------------------------------------------------------------

%------------------------------------------------------------------------------------------------------------------------------------------------------
\paragraph{Data preprocessing} We perform data preprocessing of sample cross-referencing on different union brain regions, selecting the brain regions with the required number of overlap samples. The step-by-step details of the sample cross-referencing are listed in {\bf Procedure 1}. For statistical power analysis, we select 10 union brain regions with at least $k=47$ overlap samples (See Figure~\ref{fig:0101} (a)) in subsequent experiments.

%------------------------------------------------------------------------------------------------------------------------------------------------------
\begin{algorithm}[!htbp]
  \caption*{{\bf Procedure 1.} Sample cross-referencing on a given number of brain regions.}
  \label{alg:ref}
  \footnotesize
  \begin{algorithmic}[1]
    \Require
      Datasets of 19 brain regions;
      overlap sample threshold $k$
    \Ensure
      Union regions with at least (or, at most) $k$ overlap samples
    \State Initialize $Regions=19$;
    \State Initialize the symmetric overlap sample matrix $O=\{0\}_{Regions\times Regions}$;
    \State Let $b_{i}$ and $b_{j}$ be the corresponding brain regions $i$ and $j$, $i,j=1,2,\ldots,Regions$;
    \State Denote $\#\left|b_{i}\cap b_{j}\right|$ as the overlap sample number on brain regions $i$ and $j$, $i,j=1,2,\ldots,Regions$;
     \For{each $i\in [0,Regions)$}
     	\For{each $j\in (i,Regions)$}
			\State $O[i,j]=\#\left|b_{i}\cap b_{j}\right|$;
    	\EndFor
	 \EndFor\\
    \Return Index pairs in the matrix $O$ with overlap samples $\geqslant k$ (or, $\leqslant k$)
  \end{algorithmic}
\end{algorithm}

\vspace{-1em}
\section*{Results}
\vskip -0.1in

Genome-wide gene expression datasets were obtained from the Accelerating Medicines Partnership-Alzheimer's Disease (AMP-AD) Knowledge Portal.{\footnote{See link https://www.synapse.org/\#!Synapse:syn2580853/wiki/409840}} The gene samples used in this study are part of the Mount Sinai Medical Center Brain Bank (MSBB). All samples were generated by two Affymetrix microarray platforms, Human Genome (HG) U133A and U133B, except two brain regions, amygdala (AMYG) and nucleus accumbens (NAc), by the Affymetrix HG U133 Plus 2.0 array. The gene expression data used in this study were from 19 brain cortical regions of approximately 60 individuals,{\footnote{Control samples denote the devoid of AD neuropathological changes in the brain, with $\mathrm{CDR}\leqslant 0.5$. AD samples are those with extensive AD neuropathological changes in the brain, with $\mathrm{CDR}>0.5$. See link https://www.synapse.org/\#!Synapse:syn3157699}} and the brain regions include the frontal pole (FP), occipital visual cortex (OVC), inferior temporal gyrus (ITG), middle temporal gyrus (MTG), superior temporal gyrus (STG), posterior cingulate cortex (PCC), anterior cingulate (AC), parahippocampal gyrus (PHG), temporal pole (TP), precentral gyrus (PCG), inferior frontal gyrus (IFG), dorsolateral prefrontal cortex (PFC), superior parietal lobule (SPL), prefrontal cortex (FC), caudate nucleus (CD), hippocampus (HIPP), putamen (PT), AMYG, and NAc.

%------------------------------------------------------------------------------------------------------------------------------------------------------
\begin{figure}[!htbp]
\vskip -0.14in
  \centering
    \includegraphics[width=1.0\textwidth]{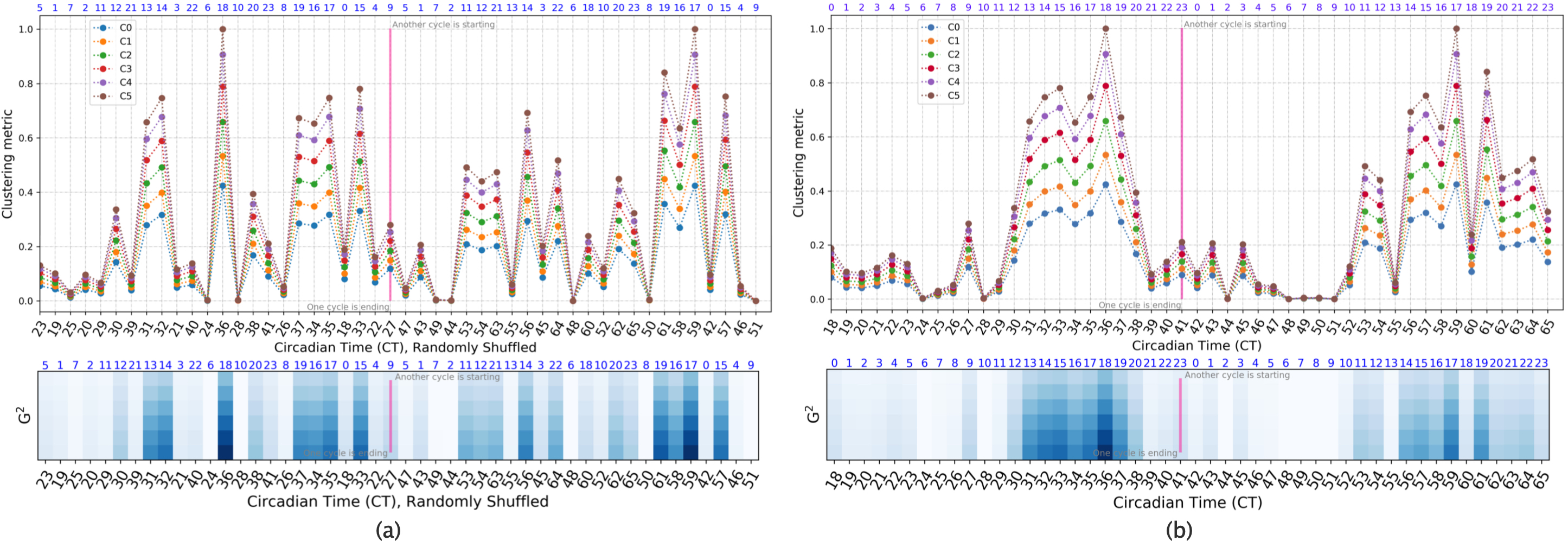}
    \vskip -0.14in
\caption{Circadian rhythmic oscillation patterns for mouse liver data. (a) With the random shuffling of time. The numbers in blue (from 0 to 23) represent the true temporal order of samples in a cycle. Because of random shuffling, these numbers are disordered; (b) With timestamps. The top panel represents the trends of circadian rhythms: 6 dotted curves show clustering metrics of 6 clusters. The bottom panel is the heat maps of $\mathbf{G}^2$, with the numbers in blue (from 0 to 23) representing the samples' order in a cycle. In (a) and (b), the vertical axis denotes the coordinates of the clusters in the latent space, and the horizontal axis denotes the circadian time given in the original data.}
\label{fig:03}
\end{figure}

In addition, we also used a time course expression dataset from mouse liver~\citep{Hughes},{\footnote{See link https://www.ncbi.nlm.nih.gov/geo/query/acc.cgi?acc=GSE11923}} which was adopted solely to validate PRIME in a cross-species and cross-organ fashion. The mouse liver samples were collected every hour for 48 hours from 3-5 mice livers. These samples were pooled and analyzed using Affymetrix microarray suite 5.0. 

%------------------------------------------------------------------------------------------------------------------------------------------------------
\paragraph{Validation}
We first randomly shuffle the mouse liver time series data in the first cycle. Then, we obtain the same shuffling of time series data in the second cycle. Subsequently, we apply PRIME to handle these disordered data and obtain Figure~\ref{fig:03} (a). It can be observed that the circadian oscillation patterns in the two different cycles resemble each other despite random shuffling. For validating the observations, we compute 3 different statistical correlations, including the Pearson, Spearman, and Kendall correlations, of circadian curves in two cycles. We obtain that Pearson correlation $correlation=0.6, p=0.002$, Spearman correlation $correlation=0.6, p=0.003$, and Kendall correlation $correlation=0.4, p=0.003$. These correlation results are substantial and demonstrate the effectiveness of PRIME on untimed data.

%------------------------------------------------------------------------------------------------------------------------------------------------------
Furthermore, PRIME does not require the data to cover the time points in the whole periodic cycle. For example, if the two cycles' data includes only the time points 7, 2, 21, 13, 14, 3, 6, 18, 10, 20, 19, 17, and 4 among the numbers in blue on the horizontal axis, PRIME can still be used to get the (partial) oscillation patterns in Figure~\ref{fig:03} (a). Subsequently, we can compute their correlation coefficients to obtain that Pearson $correlation=0.6, p=0.017$, Spearman $correlation=0.6, p=0.029$, and Kendall $correlation=0.4, p=0.038$. It is notable that PRIME effectively reveals the circadian rhythmic oscillation patterns in the absence of timestamps. 

Additionally, if the time order of samples is available, we can reorder the timestamps of data in Figure~\ref{fig:03} (a) to recover the periodic curve; See Figure~\ref{fig:03} (b). However, it is noted that recovering the periodicity curve is not essential for PRIME.

Therefore, by performing a cross-species cross-organ internal validation over the mouse liver dataset, we demonstrate that PRIME can detect circadian patterns without requiring timestamped periodic data.

\begin{figure}[!htbp]
\vskip -0.14in
  \centering
    \includegraphics[width=1\textwidth]{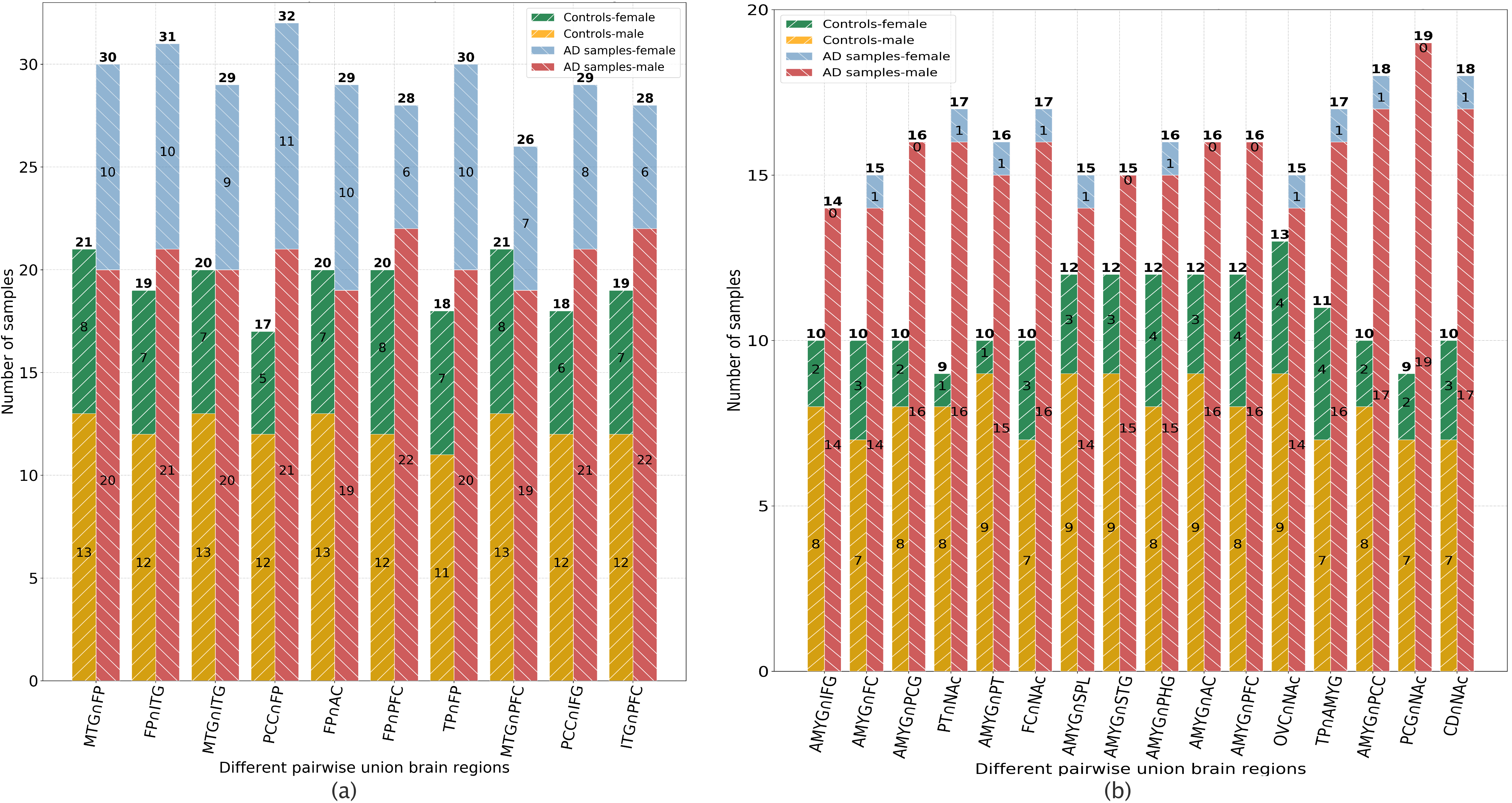}
  \vskip -0.14in
 \caption{Controls and AD patients in 10 (a) and 16 (b) union regions with overlap samples $\geqslant 47$ (a) and $\leqslant 28$ (b), respectively. The horizontal axis denotes the pairwise union brain regions, and the vertical axis represents the number of overlap samples of these union brain regions.}
 \label{fig:0101}
\end{figure}

%------------------------------------------------------------------------------------------------------------------------------------------------------
\paragraph{Analysis of untimed gene expression}

\begin{figure}[!htbp]
\vskip -0.14in
\centering
\includegraphics[width=1\textwidth]{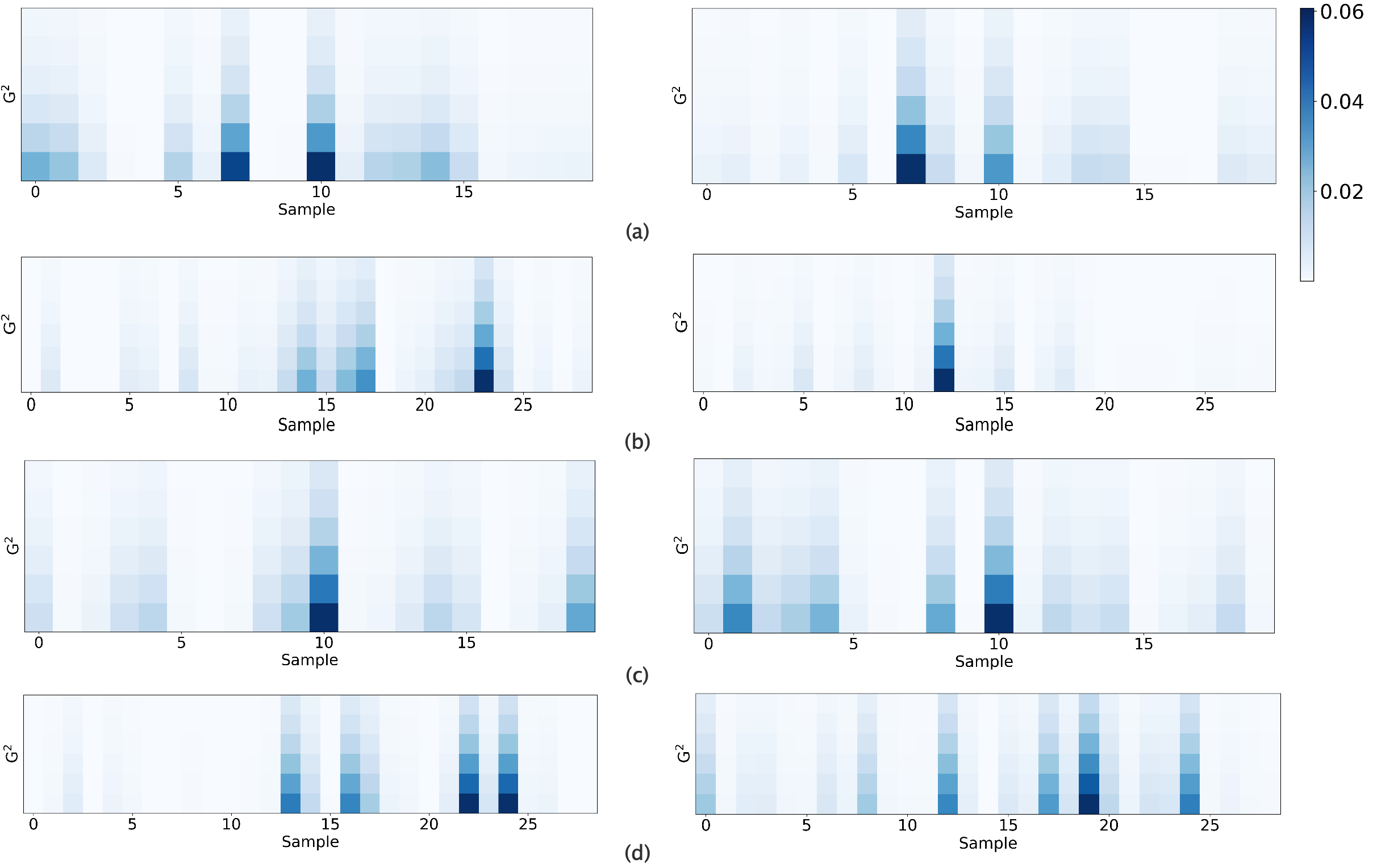}
\vskip -0.14in
\caption{Oscillation pattern’s heat maps. Genome-wide gene expression data for control (a) and AD (b); Circadian clock gene expression data for control (c) and AD (d). The left and right panels are for FP and AC, respectively.}
\label{fig:vis}
\end{figure}

Links based on genome-wide gene expression are computed for 10 union brain regions selected by sample cross-referencing with at least 47 overlap samples (See Figure~\ref{fig:0101} (a)). Further, correlations based on circadian rhythm clock genes are compared with those based on genome-wide gene expression to further verify our discoveries. From the existing studies on circadian clock genes~\citep{Jun, Cho, Morales}, we select 8 circadian clock genes, including {\sl ARNTL}, {\sl CSNK1E}, {\sl NPAS2}, {\sl NR1D2}, {\sl PER1}, {\sl PER2}, {\sl PER3}, and {\sl RORA}. We discover three clear, synchronized union brain regions of healthy controls, which are MTG$\cap$ITG, FP$\cap$AC, and FP$\cap$PFC. However, such patterns either disappear or dim for AD cases. Taking the calculation on FP$\cap$AC as an example, we visualize the heat maps of the rhythmic patterns for the genome-wide gene expression and the synthesis of the 8 circadian rhythm clock genes in Figure~\ref{fig:vis}.%------------------------------------------------------------------------------------------------------------------------------------------------------

\begin{figure}[!htbp]
\vskip -0.14in
\centering
\includegraphics[width=0.87\textwidth]{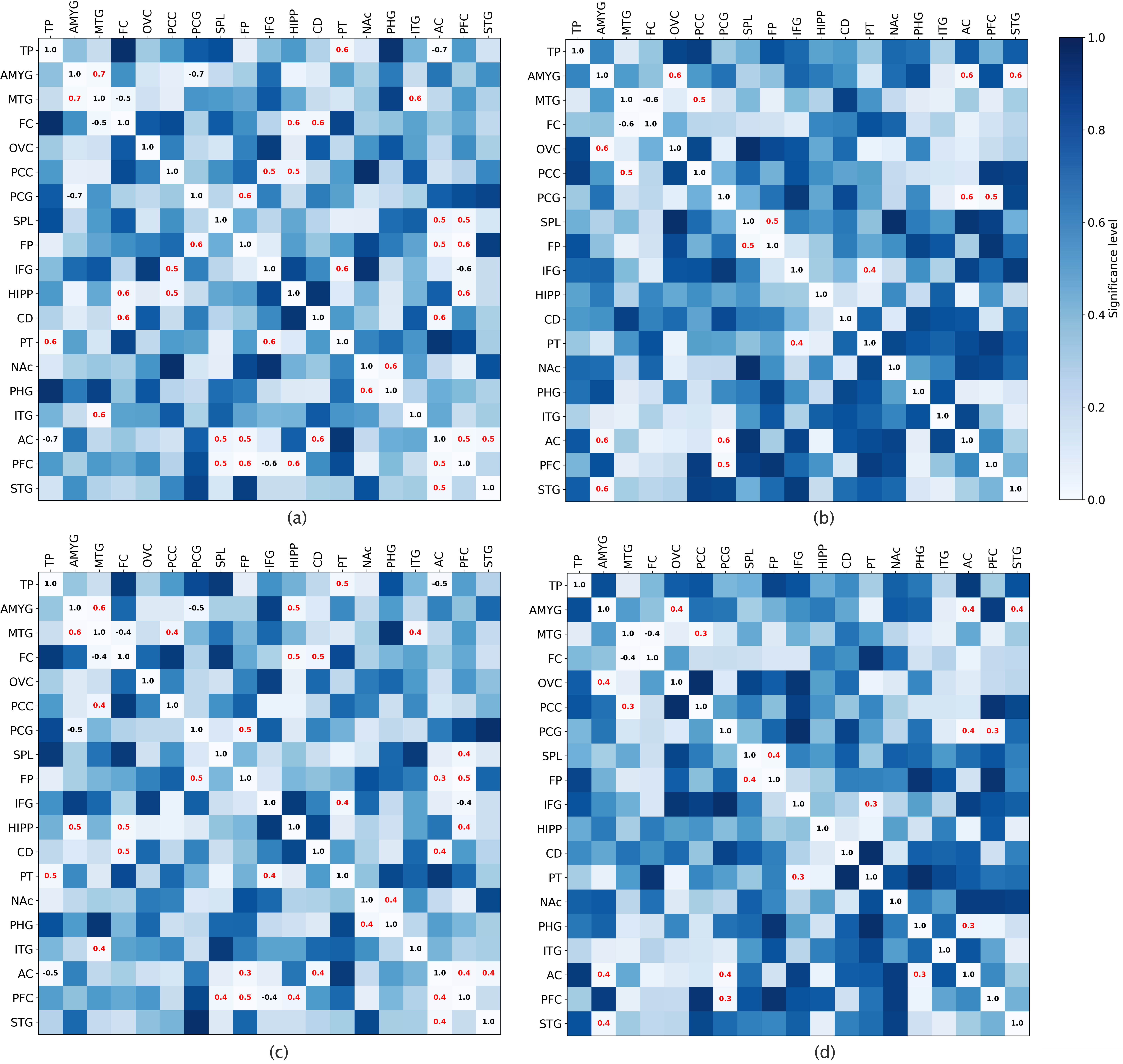}
\vskip -0.14in
\caption{Correlations of rhythmic oscillation patterns of different brain regions for control and AD. The numbers in red are statistically significant correlations. Spearman (a) and Kendall (b) correlation for control. There are 18 (a) and 19 (b) union brain regions of significant correlations. These include three clear, synchronized union brain regions, MTG$\cap$ITG, FP$\cap$AC and FP$\cap$PFC, each having at least 47 overlap samples; Spearman (c) and Kendall (d) correlation for AD. There are 8 (c) and 9 (d) union brain regions of significant correlations; The overlap samples on AC$\cap$AMYG and STG$\cap$AMYG are less than 28 (see Figure~\ref{fig:0101} (b)), and, considering the statistical significance, we ignore these brain regions in our discussion. Here, the numbers on the figures represent the correlation coefficients, while the brightness of the color bar depicts the significance.}
\label{fig:0800_01}
\end{figure}

%------------------------------------------------------------------------------------------------------------------------------------------------------
Further, pairwise correlations of 19 brain regions are computed for control and AD. We illustrate the brain regions with clear correlations in Figure~\ref{fig:0800_01}. It is evident that there exist 15 (by Spearman) and 16 (by Kendall) union brain regions showing significant correlations for control, and only 6 (by Spearman) and 7 (by Kendall) union brain regions showing significant correlations for AD cases, respectively. And for both Spearman and Kendall correlations, there are 15 overlap union brain regions of significant correlations for control, whereas only 6 overlap union brain regions for AD. Moreover, the circadian oscillation patterns of 14 union brain regions exist for control but disappear for AD. Only PT$\cap$IFG manifests correlations for control and for AD, but those for control are stronger than for AD. The correlations on the other 5 union brain regions appear to be associated with the pathology of AD.

%------------------------------------------------------------------------------------------------------------------------------------------------------

\begin{figure}[!htbp]
\vskip -0.14in
\centering
\includegraphics[width=0.98\textwidth]{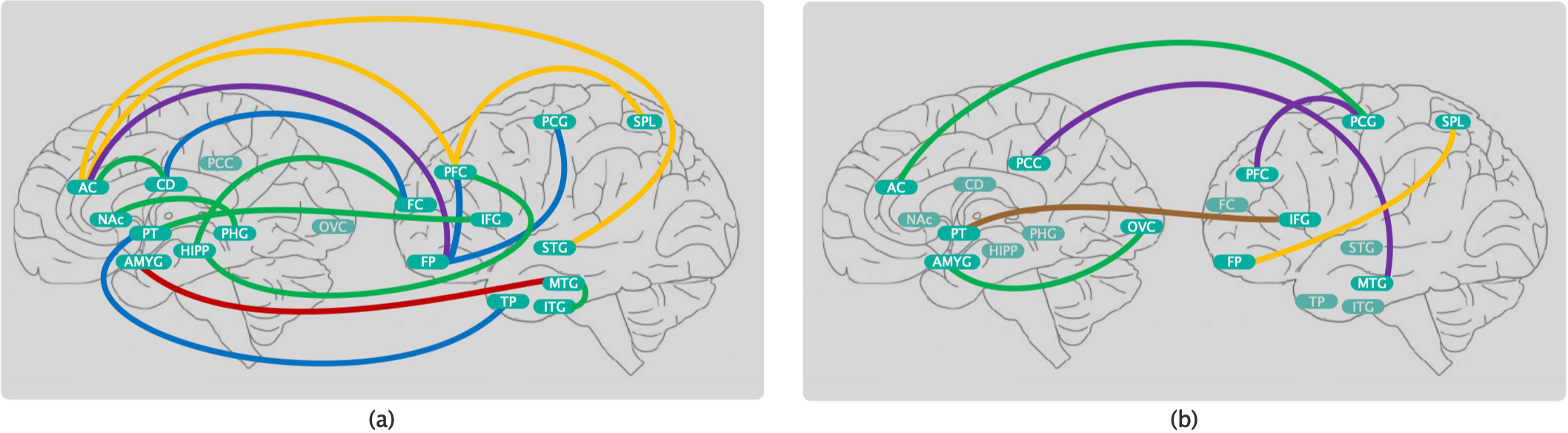}
\vskip -0.14in
\caption{Schematic representation of correlations of circadian patterns between human brain regions for control (a) and AD (b). (a) The correlation coefficients are 0.65, 0.55, 0.5, 0.45, and 0.4 for red, blue, green, orange, and purple lines, respectively. There are 15 overlap union brain regions showing substantial correlations for AD; (b) The correlation coefficients are 0.5, 0.45, 0.45, and 0.35 for green, orange, purple and brown lines, respectively. There are 6 overlap union brain regions showing positive correlations for AD. The synchronized oscillation patterns of 14 union brain regions of controls disappear for AD, while PT$\cap$IFG of control becomes weaker for AD.}
\label{fig:add02}
\end{figure}

Notably, the two overlap union brain regions are related to the hippocampus, HIPP$\cap$FC and PFC$\cap$HIPP, whereas they disappear for AD (see Figure~\ref{fig:0800_01}). Different parts of the brain are known to have specific functions; In particular, the hippocampus is pivotal for learning and memory. An emerging study has shown that auditory stimulation combined with light-induced 40-hertz gamma oscillations in the hippocampus and auditory cortex regions of the brain attenuates amyloid levels and improves cognition and memory in AD animal models~\citep{Anthony}. Our findings pinpoint the human brain regions where the inherent circadian rhythms in the brains of AD patients become out of phase or lost, thereby revealing putative human brain regions for taking a similar strategy to that for the hippocampus~\citep{Anthony} as a potential intervention or treatment for AD. In order to visualize our findings, we schematically depict the computed correlations for control and AD on the human brain regions in Figure~\ref{fig:add02}. 

In brief, our findings include 1) the consistent circadian oscillation patterns of the genome-wide profile of gene expression and the synthesis of clock genes on MTG$\cap$ITG, AC$\cap$FP, and PFC$\cap$FP for control; 2) The manifestation of more substantially correlated brain regions for control than for AD. These findings demonstrate that the synchronized oscillation patterns in many brain regions of AD cases become disarrayed compared to healthy controls. Our discovery provides new genome-wide gene expression-based evidence for the close ties of AD with circadian rhythm disorder.

\section*{Discussion}
\vskip -0.1in

%------------------------------------------------------------------------------------------------------------------------------------------------------
AD damages an individual's thoughts, memory, speech, and ability to carry out basic daily activities. While a few drugs may temporarily relieve certain mild or early cognitive symptoms of AD, they could not help permanently reduce or reverse the cognitive and functional impairment of AD. Almost all previous treatment strategies have failed in clinical trials. If AD continues to be uncontrolled, the economic burden would exceed one trillion dollars annually by 2050~\citep{Murphy}. Therefore, effective ways for early diagnosis and intervention of AD are urgently needed. 

A research thrush has recently emerged as a potential intervening means by looking beyond traditional drug strategies. It successfully uses light flickering and pulsing sounds at the frequency 40 hertz of a key brain rhythm to stimulate and reinstate the circadian rhythm of an AD affected brain region of model mice~\citep{Anthony,Rebecca,Hannah,Chinnakkaruppan,ChinnakkaruppanOther}, leading to effective amelioration of the pathological load and improvement of the cognition. A better understanding of the associations of circadian rhythms with AD would potentially benefit the development of new treatment strategies for AD. Our study has demonstrated the important role circadian rhythms play in human brain functions and provided quantitative evidence regarding how AD affects the rhythmic oscillation patterns of the human brain; thus, it should help advance our understanding of the relationship of AD with circadian rhythms of multiple brain regions. %{\footnote{Especially, this is the case for the circadian rhythms in brain regions, because disturbed circadian rhythms in the brain have deleterious consequences for the whole organism, directly or indirectly disrupting the normal functions of various organ systems~\citep{Akhilesh,Chauhan}.}}

Our approach focuses on identifying the circadian oscillation patterns in untimed data across multiple datasets and provides a new and feasible way for systematically characterizing the circadian rhythm or clock reflected on brain regions. {\sl For the first time}, correlations between multiple brain regions' circadian rhythms and AD are revealed quantitatively based on untimed genome-wide gene expression. Recently, two approaches have been developed to analyze the single clock gene’s circadian rhythm from untimed samples. \citep{Ning} has presented a statistical approach, Oscope, to extract the transcriptional dynamics of oscillating genes in single-cell gene expression data. Oscope has a prohibitively high computational complexity for genome-wide study because of the comparisons of every gene-by-gene pairing. Also, Oscope is sensitive to nonrhythmic intersubject variation. \citep{Ron} has proposed a method called cyclic ordering by periodic structure (CYCLOPS) to order periodic data without timestamps by global descriptors of expression structure. The critical step of CYCLOPS is to associate the potential temporal order with angles (uniformly) distributed in $[0,2\pi]$ by using a circular neural network. Thus, if the analyzed samples are only parts of a whole periodic cycle, as \citep{Ron} pointed out, CYCLOPS will be difficult to detect the potential oscillation pattern. It also implicitly assumes the unrealistic uniform periodic variation of circadian rhythms. Both CYCLOPS and Oscope are intended to characterize the circadian rhythms of single genes. In contrast, they are unable to analyze the orchestration of the circadian rhythmic patterns of genome-wide gene expression. However, single clock genes’ circadian rhythms can hardly be taken as a direct characterization of the circadian clocks of the brain, and the circadian rhythm of a brain region is based on coordinated regulation from all relevant genes rather than a single one. These two approaches cannot be applied to our genome-wide study.

%------------------------------------------------------------------------------------------------------------------------------------------------------
{\bf{Limitations and assumptions}}. In this study, if all samples' collection times had been the same, our approach would have become inoperative. Thus, throughout the study, our entire analysis and discussions are based on a critical assumption:{\footnote{This assumption is weak, because sample's timestamps such as time-of-death are hardly, if not impossible, all the same.}} All samples' collection times were not particularly arranged to be simultaneous; that is, the samples' timestamps were not all the same.

%------------------------------------------------------------------------------------------------------------------------------------------------------
\section*{Conclusions}
\vskip -0.1in

The disruption of brain circadian rhythm is a crucial symptom for patients with AD. Current studies and accumulating clinic evidence suggest that such a symptom usually occurs in the early phase of AD and probably precedes the development of cognitive impairment. However, the full and systematic circadian rhythm orchestration of genome-wide gene expression in the human brain and its potential dysregulation in AD remains unknown. This study explores the relationships between circadian rhythms of different brain regions and AD with untimed genome-wide gene expression data. To this end, we developed an unsupervised machine learning-based approach, PRIME, to detect and analyze oscillation patterns in untimed high-dimensional data across multiple brain regions. This innovative approach is able to discover potential circadian oscillation patterns from untimed high-dimensional expression data across multiple datasets. It does not need the data from an entire periodic cycle, nor does it need to impose a high complexity for high-dimensional gene expression data. It mainly leverages unsupervised learning and thus does not need any timestamps or labels that may be tedious to obtain. By PRIME, we reveal clear, synchronized oscillation patterns in 15 pairs of brain regions of control, whereas these oscillation patterns either disappear or dim for AD. 

\subparagraph{Acknowledgments}
\vskip -0.1in
This work was partially supported by the NIH grants R21AG070909, R56NS117587, R01HD101508, and ARO W911NF-17-1-0040. 

The results published here are in whole or in part based on data obtained from the AD Knowledge Portal. These data were provided by the Rush Alzheimer’s Disease Center, Rush University Medical Center, Chicago.
\vskip -0.1in

% References as numbers
\makeatletter
\renewcommand{\@biblabel}[1]{\hfill #1.}
\makeatother

\vskip -0.2in
% unstr is used to keep citation order
\begin{center}
\bibliographystyle{vancouver}
\bibliography{references}  
\end{center}

\end{document}